# Mô hình hóa số bàn thắng mỗi trận đấu của đội chủ nhà tại giải Ngoại Hạng Anh bằng phương pháp hồi quy tuyến tính tổng quát với biến phụ thuộc có phân phối Poisson

Lê Thanh Tùng    Phạm Tuấn Cường

*Lớp Cao học Khoa học Dữ liệu khóa 3, Khoa Toán - Cơ - Tin,*
*Đại học Khoa học Tự nhiên, Đại học Quốc gia, Hà Nội, Việt Nam*

**Tóm tắt**

Giải ngoại hạng Anh được biết đến là 1 trong những giải đấu hấp dẫn nhất và một trong những yếu tố được quan tâm nhất là số bàn thắng các đội bóng ghi được tại đây. Trong bài viết này, tác giả hướng đến việc phân tích dữ liệu và mô hình hóa số bàn thắng mà một đội bóng ghi được trên sân nhà tại giải đấu này.

Nghiên cứu sẽ cố gắng để xác định xem phải chăng số bàn thắng các đội tuyển ghi được trên sân nhà có phân phối Poisson và liệu có mối liên hệ nào giữa số bàn thắng với các yếu tố như số thẻ đỏ, số thẻ vàng, số quả phạt góc đội được nhận cũng như chính bản thân đội bóng đó (đội mạnh hoặc yếu thì số bàn ghi được cũng sẽ khác nhau).

Kết quả thu được cho thấy hoàn toàn có thể dùng phân phối Poisson để đặc trưng cho số bàn thắng 1 đội ghi được trên sân nhà cũng như cũng có mối tương quan đáng kể giữa các biến phụ thuộc như số thẻ, số phạt góc đến số bàn thắng của đội.

## 1. Giới thiệu:

Trong quá khứ đã có một số nghiên cứu liên quan đến số bàn thắng ghi được trong bóng đá.

Trong nghiên cứu "Goal scoring patterns in major European soccer leagues" [1], các tác giả Giampietro Alberti và F. Marcello Iaia đã cố gắng tìm ra phân bố của số bàn thắng ghi được trong một trận đấu dựa theo thời gian thi đấu. Với giả thuyết là số bàn thắng ghi được ở hiệp 2 của trận đấu là lớn hơn hiệp 1, cũng như số bàn thắng ghi được trong khoảng 15 phút cuối trận sẽ cao hơn những khoảng thời gian khác trong trận đấu. Phương pháp mà các tác giả sử dụng trong nghiên cứu này là phân tích thống kê Nonparametric Chi square để kiểm định giả thuyết về sự khác biệt trong số bàn thắng ghi được giữa 2 hiệp đấu và giữa khoảng thời gian 15 phút cuối trận với các khoảng 15 phút khác (so sánh khoảng 75–90 phút với các khoảng 0–15, 15–30, 30–45, 45–60, 60–75 phút).

Nghiên cứu trên sử dụng dữ liệu là tỷ số các trận đấu ở các giải bóng đá hàng đầu châu Âu như ngoại hạng Anh, Series A và Ligue 1. Đồng thời cũng đã chỉ ra được rằng hiệp 2 có số bàn thắng ghi được nhiều hơn so với hiệp 1 và khoảng thời gian 15 phút cuối trận là khoảng thời gian ghi bàn nhiều nhất. Tuy nhiên, nghiên cứu chưa chỉ ra được những yếu tố có ảnh hưởng đến số lượng bàn thắng này.

Trong một nghiên cứu khác về số bàn thắng ghi được ở các trận đấu thuộc Premier League,

Modeling Scores in the Premier League: Is Manchester United Really the Best? [2], tác giả Alan J. Lee muốn xác định xác suất để một đội bóng giành chiến thắng (ghi được nhiều bàn thắng hơn) tại Premier League, với dữ liệu từ 380 trận đấu trong mùa giải 95/96.

Trong nghiên cứu này, tác giả giả thuyết rằng số bàn thắng ghi được của một đội là tuân theo phân phối Poisson và từ đó xây dựng mô hình GLM mới biến phụ thuộc là số bàn thắng còn biến giải thích là các đội bóng tham gia thi đấu trong trận đấu đó. Đây là một mô hình khá đơn giản và có thể sử dụng để xác định xác suất giành chiến thắng của 1 đội trong 1 trận đấu xác định (thông qua việc ước lượng số bàn thắng đội đó ghi được so với đối thủ).

Tuy nhiên, mô hình này chưa kiểm định được giả thuyết: số bàn thắng ghi được của 1 đội tuân theo phân phối Poisson cũng như chưa đánh giá đầy đủ các yếu tố ảnh hưởng đến số bàn thắng ghi được (ví dụ số thẻ vàng, thẻ đỏ, số pha sút phạt, số quả phạt góc,…).

Ngoài ra, dữ liệu mà nghiên cứu sử dụng cũng chưa đủ lớn (đang giới hạn số bàn thắng ở mùa giải 96/96).

Nghiên cứu "Studying some aspects of data from the professional Dutch soccer competition" [3] của tác giả Dieuwe van Bergen en Henegouwen cũng đề xuất sử dụng mô hình GLM Poisson Regression để tìm ra mối liên hệ giữa sự hiệu số bàn thắng của 2 đội bóng trong 1 trận đấu với những yếu tố khác như loại sân (cỏ nhân tạo hay cỏ thật).

Nghiên cứu này có biến phụ thuộc là hiệu số bàn thắng và biến này có thể có giá trị âm, dương hoặc bằng 0. Biến này cũng có phân phối chuẩn. Tuy nhiên, do giá trị hiệu số bàn thắng này là 1 giá trị rời rạc (các số nguyên như -2, -1, 0, 1, 2, …) nên biến phụ thuộc này có thể được xem là phân phối Skellam (hiệu của 2 biến ngẫu nhiên có phân phối Poisson).

Sau đó, tác giả có sử dụng Poisson regression để xác định gần đúng phụ thuộc của biến hiệu số bàn thắng với những biến giải thích khác như chất lượng sân bóng. Do biến ngẫu nhiên phân phối Skellam chỉ gần đúng có thể sử dụng poisson regression nên mô hình mô tả chưa đủ chính xác được về số bàn thắng các đội ghi được.

Như vậy có thể thấy, các mô hình đã có hoặc là chưa đánh giá được đầy đủ ảnh hưởng của các biến giải thích vào biến phụ thuộc "số bàn thắng 1 đội ghi được trong 1 trận đấu" hoặc là chưa có đầy đủ các bước phân tích GLM cần có như chọn biến, chẩn đoán mô hình. Các mô hình còn sơ sài và việc chọn biến mang tính cảm tính, thiếu cơ sở toán học.

Vì vậy, nhóm quyết định sẽ phân tích để tìm ra mối quan hệ của các biến giải thích như số thẻ, số phạt góc, đội sân nhà, đội sân khách, … đối với số bàn thắng mà đội chủ nhà ghi được trong 1 trận đấu tại Premier League.

Với việc sử dụng mô hình hồi quy Poisson với đầy đủ các bước từ xác định phân phối của biến ngẫu nhiên, chọn biến, đến chẩn đoán mô hình.

## 2. Trình bày phương pháp

Do ta chỉ quan tâm đến biến phụ thuộc là Số bàn thắng đội nhà ghi được (cả trận) - FTHG nên ta chỉ xét những biến ảnh hưởng trực tiếp nhất đến kết quả của trận đấu (các biến trên). Ngoài ra, các biến của đội khách phần lớn sẽ không được xét (ví dụ số phạt góc, số cú sút trúng đích của đội khách.

### 2.1. Nhập liệu và mô tả dữ liệu

Dữ liệu được sử dụng trong nghiên cứu được lấy từ trang web football-data.co.uk [4], bao gồm dữ liệu của tất cả các trận đấu trong khuôn khổ giải Ngoại Hạng Anh từ mùa giải 2000/2001 đến 2020/2021.

Do nghiên cứu muốn tập trung vào khả năng ghi bàn của đội chủ nhà nên một số dữ liệu được sử dụng là:

- Date: Ngày trận đấu diễn ra
- HomeTeam: Đội chủ nhà
- FTHG: Số bàn thắng đội nhà ghi được (cả trận)

- Attendance: Số khán giả có mặt tại sân
- Referee: Trọng tài điều khiển trận đấu
- HTAG: Số bàn đội khách ghi được trong hiệp 1
- HST: Số cú sút trúng đích của đội chủ nhà
- HHW: Số cú sút trúng cột dọc và xà ngang của đội chủ nhà
- HC: Số quả phạt góc của đội chủ nhà
- HO: Số lần việt vị của đội chủ nhà
- HR: Số thẻ đỏ của đội chủ nhà
- AR: Số thẻ đỏ của đội khách

*2.1.1. Kiểm tra các biến bị thiếu dữ liệu*

Ta sử dụng hàm vis_dat để vẽ biểu đồ về số lượng quan sát đối với từng biến trong tập dữ liệu, dễ thấy hầu hết các biến (các cột) trong dữ liệu đều không bị thiếu.

Chỉ có biến **Attendance, HHW, HO** là bị thiếu khá nhiều (do những năm gần đây nguồn data không có thu thập dữ liệu về những thông số này). Phần màu xám trong đồ thị thể hiện các biến bị thiếu (hình 1). Do đó, ta sẽ bỏ 3 cột này ra khỏi tập dữ liệu của mình.

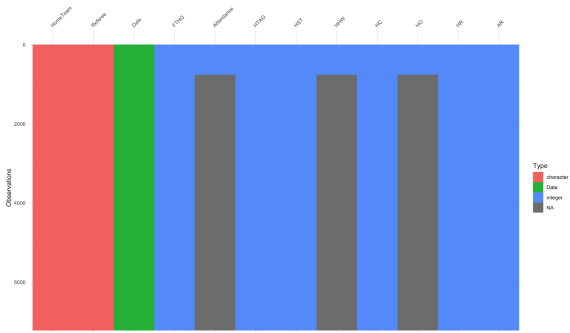

*Hình 1: Biểu đồ mô tả các biến Attendance, HHW, HO có hiện tượng thiếu dữ liệu.*

*2.1.2. Dữ liệu về trọng tài (Referee)*

Trọng tài có thể được xem là 1 yếu tố ảnh hưởng đến trận đấu bóng đá, tuy nhiên, đây là 1 biến dạng category nên nếu có quá nhiều giá trị sẽ khiến mô hình thiếu chính xác.

Số lượng trọng tài là 159, trong khi chỉ có hơn 7000 trận đấu, do đó ta cũng sẽ loại biến này ra khỏi mô hình. Như vậy, nghiên cứu sẽ chỉ giữ các biến Date, HomeTeam, FTHG, HTAG, HST, HC, HR, AR trong dữ liệu vì đây là những biến có thể ảnh hưởng nhiều nhất đến kết quả trận đấu. Trong đó, Date sẽ không được xét trong model mà chỉ dùng để phân biệt giữa các trận đấu.

*2.1.3. Biểu đồ số bàn thắng ghi được (FTHG)*

Do biến ngẫu nhiên mà ta quan tâm là số bàn thắng đội nhà ghi được, ta sẽ vẽ biểu đồ cột về số bàn thắng này (hình 2).

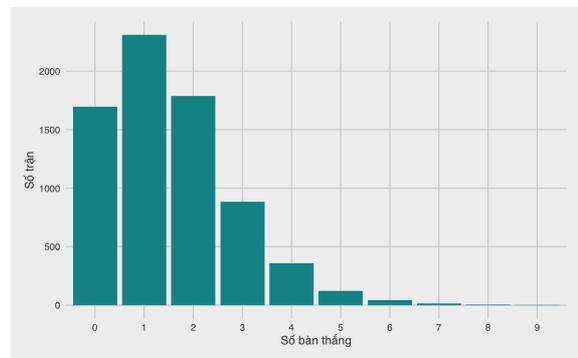

*Hình 2: biểu đồ mô tả số bàn thắng ghi được của đội nhà (FTHG).*

Từ biểu đồ, có thể thấy phần lớn các trận đấu đều kết thúc với 1 bàn thắng dành cho đội chủ nhà. Sau đó số bàn thắng giảm dần, khi mà có rất ít trận, các đội chủ nhà ghi được trên 5 bàn thắng.

Có thể suy đoán, số bàn thắng đội nhà ghi được là 1 biến ngẫu nhiên có phân phối Poisson.

Một số đặc trưng của biến FTHG (hình 3):

```
##   min Q1 median Q3 max     mean       sd    n missing
##     0  1      1  2   9 1.522438 1.299001 7220       0
```

*Hình 2: Thông số thống kê mô tả biến số bàn thắng ghi được của đội nhà (FTHG).*

Như vậy, số bàn thắng tối đa 1 đội chủ nhà ghi được trong 1 trận đấu là 9, tối thiểu là 0 và trung bình là 1.52 bàn thắng.

Để xác định phân phối của biến FTHG, cần kiểm định Chi-square.

*2.1.4. Biểu đồ các biến giải thích*

Tương tự với biến phụ thuộc "số bàn thắng đội nhà ghi được", các biến giải thích cũng là biến rời rạc.

Biến AR, HR là thẻ đỏ của đội nhà và đội khác, do đặc thù của bóng đá, phần lớn các trận đấu sẽ không có thẻ đỏ, vì thế có thể thấy phân bố giá trị biến này tập trung ở giá trị 0.

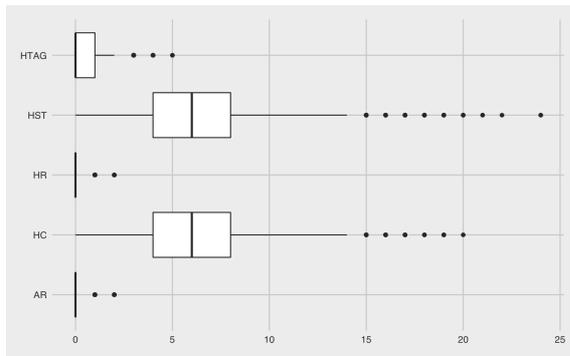

*Hình 3: Biểu đồ Boxplot mô tả các biến HTAG, HST, HR, HC, AR*

HTAG là số bàn thắng đội khách ghi được trong hiệp 1, vì thời gian hiệp 1 ngắn, nên thường các đội không ghi được bàn nào trong hiệp này. Do đó, biến này cũng phân bố phần lớn ở giá trị 0.

Như vậy, 3 biến AR, HR và HTAG có khá nhiều giá trị 0.

HC là số phạt góc cùng HST là số cú sút trúng đích của đội có phân phối tương tự nhau, trong phần lớn các trận đấu đội chủ nhà đều có từ 4 đến 8 quả phạt góc hoặc cú sút trúng đích.

Các biểu đồ boxplot của biến HC và biến HST cho thấy phân bố tần suất có xu hướng lệch phải. Do đó để tăng tính giải thích, nhóm nghiên cứu sử dụng kĩ thuật logarith để đạt được dữ liệu mới có phân phối sát với phân phối chuẩn hơn (hình 4).

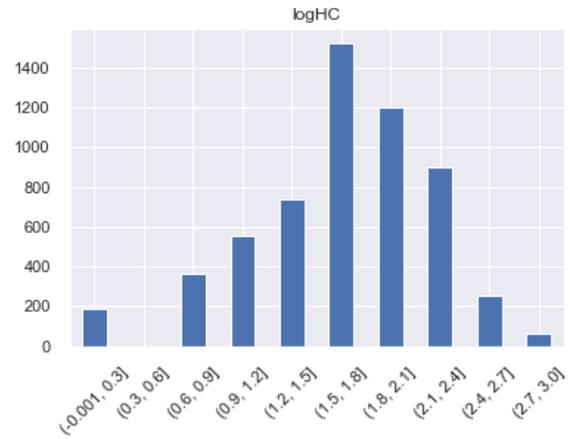

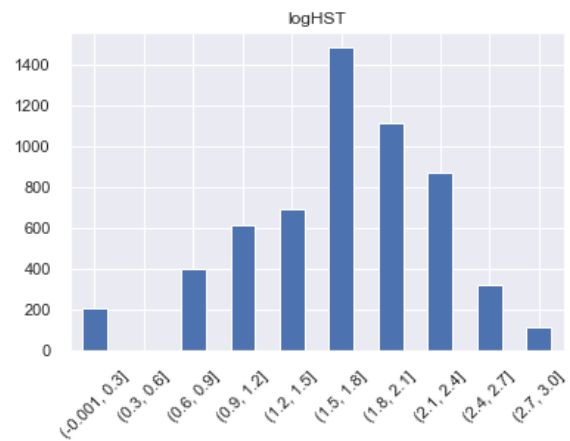

*Hình 4: Biểu đồ cột mô tả các biến HST, HC sau khi được biến đổi logarith*

*2.1.5. Biến các đội chủ nhà (HomeTeam):*

Sau khi thống kê, ta có 43 đội tham gia giải ngoại hạng trong giai đoạn từ năm 2000 đến 2021

Thống kê số trận đấu của các đội, có thể thấy có những đội bóng mạnh, góp mặt thường xuyên ở Ngoại hạng Anh sẽ có số trận đấu lớn như Arsenal, Manchester United, … với 361 trận mỗi đội. Những đội bóng nhỏ như Coventry, Ipswich, … chỉ có 19 trận đấu (hình 5).

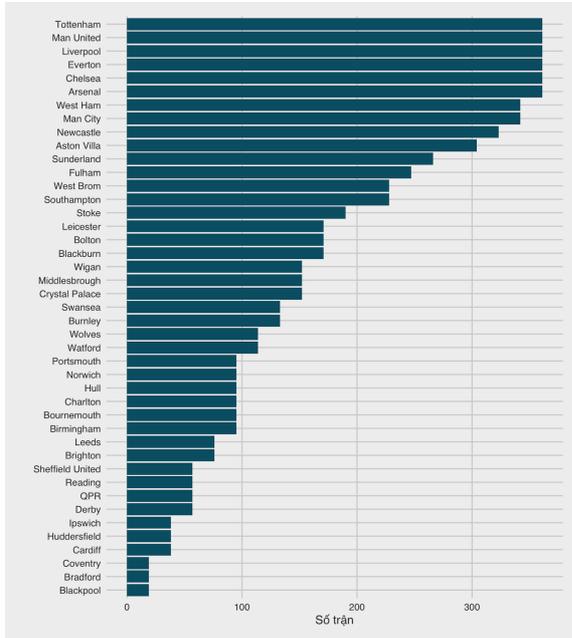

*Hình 5: Số lượng trận đấu mỗi đội tham gia*

### 2.2 Kiểm tra phân phối của biến phụ thuộc - số bàn thắng của đội nhà (FTHG):

Do có rất ít đội ghi được nhiều hơn 5 bàn, nên ta sẽ tính tất cả những trận có trên 5 bàn sẽ có cùng giá trị đối với biến FTHG là: > 5 goals

Trong đó, FTHG là số bàn thắng FTHG, ActualMatches là số trận ghi được số bàn thắng đó (hình 6).

```
##   FTHG         ActualMatches
##   <chr>                <int>
## 1 0                     1695
## 2 1                     2310
## 3 2                     1787
## 4 3                      885
## 5 4                      357
## 6 5                      122
## 7 more than 5             64
```

*Hình 6: Số trận đấu thực tế (với số bàn tương ứng)*

Ta đã thấy số bàn thắng của tất cả các đội có phân phối gần giống Poisson, và để kiểm tra trước hết ta sẽ tạo ra mẫu có phân bố Poisson với giá trị trung bình là giá trị trung bình của biến FTHG.

Ghép giá trị của PoisProb vào cột trong summary_data để so sánh với ActualMatches (hình 7).

```
##        FTHG ActualMatches PoisProb ExpectedMatches
## 1         0          1695    0.218            1574
## 2         1          2310    0.332            2397
## 3         2          1787    0.253            1827
## 4         3           885    0.128             924
## 5         4           357    0.049             354
## 6         5           122    0.015             108
## 7 more than 5          64    0.005              36
```

*Hình 7: Số trận đấu thực tế (với số bàn tương ứng) so với số trận tính theo phân phối Poisson*

Có thể thấy, giá trị thực tế của số lần xuất hiện 1 giá trị của biến FTHG là ActualMatches khá sát với giá trị kỳ vọng ExpectedMatches (hình 8).

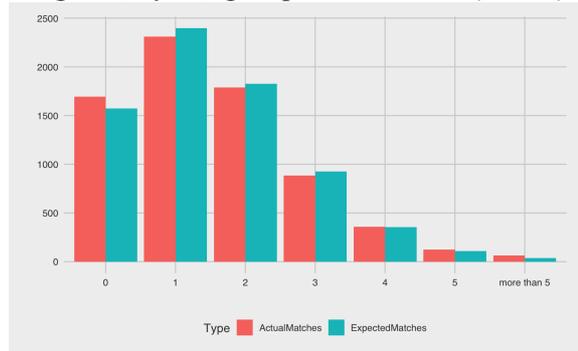

*Hình 8: Đồ thị giữa số trận thực tế (với số bàn tương ứng) với số trận tính theo phân phối Poisson*

Tuy nhiên, nếu kiểm định Chi-square với giả thuyết

*H0: Phân phối của FTHG là phân phối Poisson* và

*H1: Phân phối của FTHG không phải là phân phối Poisson*

thì ta được kết quả như hình 9:

```
chi_result <- chisq.test(poisson_table$ActualMatches, p = poisson_table$PoisProb)
chi_result

##
##  Chi-squared test for given probabilities
##
## data:  poisson_table$ActualMatches
## X-squared = 38.314, df = 6, p-value = 0.0000009752
```

*Hình 9: Kiểm định Chi-squared với H0 phân phối của ActualMatches là Poisson*

Giá trị p-value nhỏ hơn 0.05 nên với mức ý nghĩa alpha là 5% ta sẽ bác bỏ giả thuyết H0, nói cách khác, không thể kết luận phân phối này của FTHG với tất cả các trận và tất cả các đội là phân phối Poisson.

### 2.3 Lựa chọn đội

Do một số đội bóng có quá ít trận đấu hoặc ghi quá nhiều bàn thắng trong 1 trận (số lượng trận đấu có hơn 5 bàn rất nhiều) nên dữ liệu tổng của tất cả các đội sẽ không có phân phối Poisson. Vì thế, ta sẽ chọn các đội có số bàn thắng hợp lý hơn để mô hình có thể hoạt động chính xác (hình 10).

```
for (name in team_name$HomeTeam){
  chi_result <- check_poisson(name = name)
  p_value <- ifelse(typeof(chi_result) == "double", 0, chi_result$p.value)
  cat(name, " has p-value from chisq test: ", p_value, "\n")
}

## Charlton  has p-value from chisq test:  0
## Chelsea  has p-value from chisq test:  0.1622334
## Coventry  has p-value from chisq test:  0
## Derby  has p-value from chisq test:  0
## Leeds  has p-value from chisq test:  0.8694502
## Leicester  has p-value from chisq test:  0
## Liverpool  has p-value from chisq test:  0.6594176
## Sunderland  has p-value from chisq test:  0
## Tottenham  has p-value from chisq test:  0.8094301
## Man United  has p-value from chisq test:  0.666615
## Arsenal  has p-value from chisq test:  0.970537
## Bradford  has p-value from chisq test:  0
```

*Hình 10: Kiểm định Chi-squared với từng đội*

Như vậy, loại trừ các đội bóng có số bàn thắng quá ít (như Cardiff) hoặc quá nhiều (như Man City), ta sẽ còn lại các đội có số bàn thắng tuân theo phân phối Poisson

```
## [1] "Chelsea"
## [1] "Leeds"
## [1] "Liverpool"
## [1] "Tottenham"
## [1] "Man United"
## [1] "Arsenal"
## [1] "Middlesbrough"
## [1] "Everton"
## [1] "Newcastle"
## [1] "Aston Villa"
## [1] "Blackburn"
## [1] "Fulham"
```

*Hình 11: Các đội được chọn*

### 2.4. Tìm kiếm mô hình

Để thuận tiện cho việc xây dựng mô hình hồi quy tuyến tính cho tất cả các trường hợp, nhóm nghiên cứu sử dụng ngôn ngữ Python với các thư viện: numpy, pandas (tổ chức, biến đổi bảng dữ liệu), itertools (lập tổ hợp các bộ biến giải thích), stats (tính toán xác suất thống kê), statsmodels (hồi quy tuyến tính tổng quát), matplotlib, seaborn (visualize các biểu đồ chẩn đoán.)

Để giải thích sự phụ thuộc của "số bàn thắng mỗi trận đấu của đội chủ nhà" (FTHG) thông qua các biến giải thích HomeTeam, FTHG, HTAG, logHST, logHC, HR, AR; nhóm nghiên cứu sử dụng mô hình hồi quy tuyến tính tổng quát (Generalized Linear Model - GLM) với biến phụ thuộc có phân phối Poisson.

Để lựa chọn mô hình phù hợp, nhóm nghiên cứu quyết định xây dựng mô hình hồi quy cho từng trường hợp tổ hợp của biến giải thích. Cụ thể chúng ta có 6 biến giải thích, như vậy tổng số mô hình có thể xây dựng là:

$$SốTrườngHợp = \sum_{i=1}^{6} \binom{6}{i} = 63$$

Áp dụng phương thức smf.glm với thông số family=sm.families.Poisson() cho lần lượt từng trường hợp tổ hợp của biến giải thích, nhóm nghiên cứu thu được danh sách các mô hình với thông số về Deviance, pearson_chi2, llf(log likelihood function), df_resid (bậc tự do của phần dư) và AIC tương ứng như bảng 1.

Với các mô hình thu được, nhóm nghiên cứu tiếp tục kiểm định sự phù hợp của mô hình với kiểm định GoodnessOfFit. cụ thể, nhóm nghiên cứu tính toán giá trị một trừ đi xác suất tích lũy theo phân phối Chi bình phương với các thông số Deviance và bậc tự do của phần dư. Áp dụng trong Python, việc tính toán xác suất này sử dụng thư viện stats:

*1 - stats.chi2.cdf( row["deviance"], row["df_resid"])*

Với các giả thiết: *H0: mô hình không phù hợp, H1: mô hình phù hợp* và độ tin cậy 5% nhóm nghiên cứu nhận thấy rằng các mô hình có xác suất Chi bình phương nhỏ hơn 0.05 là không phù hợp và nên bị loại bỏ. Như vậy sau khi kiểm định sự phù hợp của mô hình với kiểm định GoodnessOfFit, số mô hình phù hợp giảm xuống chỉ còn 31 mô hình và được sắp xếp theo thứ tự chỉ số AIC tăng dần như bảng 2.

Để lựa chọn mô hình phù hợp nhất, nhóm nghiên cứu sử dụng chỉ số AIC (Akaike information criterion). Với một nhóm các mô hình thống kê được ước lượng dựa trên cùng một tập dữ liệu, chỉ số AIC ước lượng chất lượng của mô hình thống kê và sử dụng kết quả này để đánh giá mô hình tốt nhất. Chỉ số AIC được tính như sau : $AIC = 2k - 2\ln(\hat{L})$

Như vậy có thể thấy để phản ánh chất lượng của mô hình, AIC sử dụng số biến giải thích (k) và logarithh của log-likelihood. AIC càng nhỏ thì mô hình càng phù hợp.

Thông qua AIC, nhóm nghiên cứu kết luận rằng mô hình [1]: *FTHG ~ HTAG + logHST + logHC + HR + AR + HomeTeam* là mô hình phù hợp nhất. Các thông số của mô hình được thể hiện ở bảng 3, bảng 4.

| Dep. Variable: | FTHG | No. Observations: | 5852 |
|---|---|---|---|
| Model: | GLM | Df Residuals: | 5821 |
| Model Family: | Poisson | Df Model: | 30 |
| Link Function: | log | Scale: | 1 |
| Method: | IRLS | Log-Likelihood: | -8472.2 |
| No. Iterations: | 5 | Deviance: | 5622.5 |
| Covariance Type: | nonrobust | Pearson chi2: | 4.71E+03 |

*Bảng 3: các thông số của mô hình [1]*

### 2.5. Chẩn đoán mô hình

Để chẩn đoán mô hình, nhóm nghiên cứu tiến hành vẽ các biểu đồ phần dư (pearson residual) so với giá trị biến phụ thuộc tính được sau mô

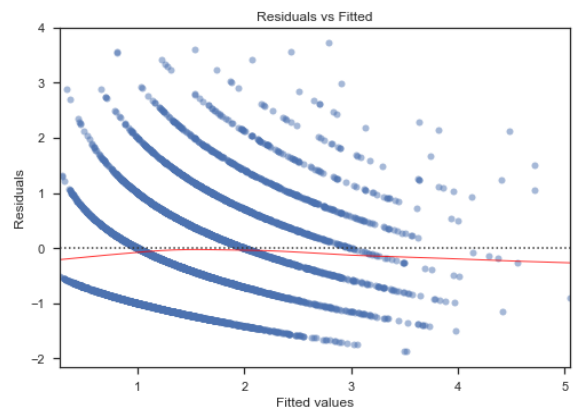

*Hình 12: biểu đồ phần dư so với giá trị biến phụ thuộc tính được sau mô hình hóa .*

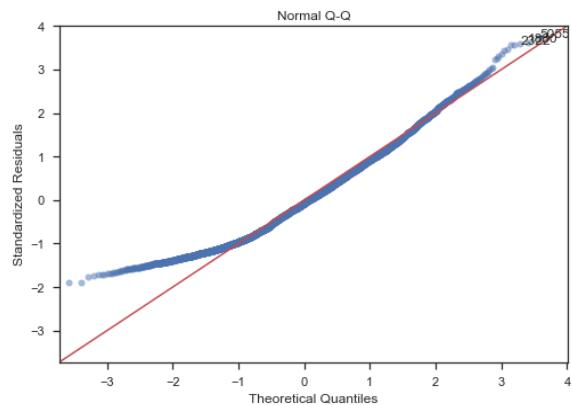

*Hình 13: đồ thị Q-Q*

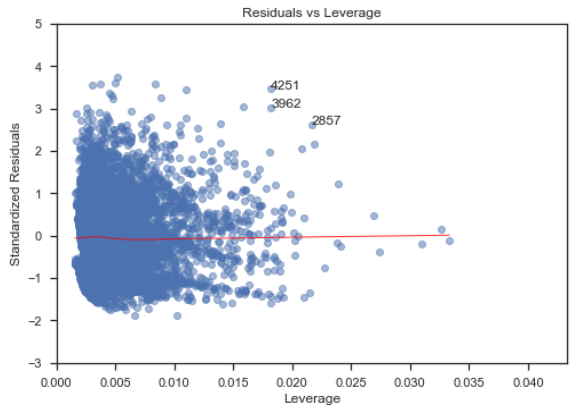

*Hình 14: Biểu đồ phần dư chuẩn hóa so với Leverage.*

hình hóa (Fitted Values); Biểu đồ Q-Q, và Biểu đồ phần dư chuẩn hóa (standardized Residuals) so với Leverage. Các kết quả thu được như hình 12, 13, 14.

Biểu đồ 13 cho thấy phần lớn các giá trị phần dư đã có phân phối chuẩn. Biểu đồ 11 cho thấy các giá trị phần dư nằm phân tán đều, do đó không có các lỗi phương sai thay đổi. Biểu đồ 14 cho thấy không thực sự có các giá trị ngoại vi đáng kể.

Mặc dù vậy căn cứ trên mô hình, nhóm nghiên cứu vẫn tiến hành loại bỏ một số quan sát mà các đồ thị cho là ngoại vi như: 3962, 4251, 2857, 5055, 1880, 2322. Sau khi bỏ bớt các quan sát, các thông số của mô hình [1] sau điều chỉnh được thể hiện ở bảng 5 và bảng 6.

| Dep. Variable: | FTHG | No. Observations: | 5846 |
|---|---|---|---|
| Model: | GLM | Df Residuals: | 5815 |
| Model Family: | Poisson | Df Model: | 30 |
| Link Function: | log | Scale: | 1 |
| Method: | IRLS | Log-Likelihood: | -8465.3 |
| No. Iterations: | 5 | Deviance: | 5613.8 |
| Covariance Type: | nonrobust | Pearson chi2: | 4.70E+03 |

*Bảng 5: các thông số của mô hình [1] sau điều khi bỏ bớt quan sát.*

Sau khi tiến hành bỏ bớt quan sát, nhóm nghiên cứu nhận thấy các quan sát theo đánh giá ban đầu là không trọng yếu, đã thực sự không có tác động đáng kể đến mô hình [1]. Do đó, nhóm kết luận mô hình [1] đã có thể giải thích cho mối liên hệ giữa biến giải thích và các biến độc lập.

### 3. Kết luận

Theo mô hình [1], số bàn thắng đội chủ nhà ghi được trong 1 trận đấu phụ thuộc vào số bàn thắng đội khách ghi được trong hiệp 1, logarithh của số cú sút trúng đích của đội chủ nhà, logarithh của số quả phạt góc đội chủ nhà thực hiện, số thẻ đỏ của 2 đội, và bản thân đội chủ nhà.

Có thể thấy các đội bóng mạnh như Chelsea, Manchester United có hệ số trong mô hình lớn hơn các đội yếu, do đó khả năng để các đội này ghi nhiều bàn thắng hơn cũng lớn hơn.

Số bàn thắng đội khách ghi được trong hiệp 1 cũng là 1 yếu tố ảnh hưởng nhưng hệ số khá nhỏ (gần sát với 0).

Có 2 yếu tố ảnh hưởng mạnh nhất đến số bàn thắng đội chủ nhà ghi được là số cú sút trúng đích và số thẻ đỏ của đội khách. Điều này có thể được giải thích là đội sút trúng khung thành nhiều hơn sẽ có khả năng ghi bàn tốt hơn. Đồng thời, việc đối thủ bị thẻ đỏ tạo ra lợi thế lớn để mang về nhiều bàn thắng hơn.

Thông qua mô hình, chúng ta xác định được các yếu tố ảnh hưởng đến số bàn thắng 1 đội ghi được cũng như mức độ ảnh hưởng mạnh yếu của từng yếu tố. Nhờ đó giúp các đội có thể có chiến thuật tiếp cận trận đấu tốt hơn.

*Bảng 1: Kết quả hồi quy tuyến tính tổng quát của tất cả các mô hình có thể tìm được*

| | model | deviance | pearson_chi2 | llf | df_resid | AIC |
|---|---|---|---|---|---|---|
| 0 | HTAG | 7234.792327 | 6390.101744 | -9278.392 | 5850 | 18560.784 |
| 1 | logHST | 6222.33253 | 5193.06051 | -8772.1621 | 5850 | 17548.3242 |
| 2 | logHC | 7241.81159 | 6392.376432 | -9281.90163 | 5850 | 18567.8033 |
| 3 | HR | 7206.607732 | 6349.947591 | -9264.29971 | 5850 | 18532.5994 |
| 4 | AR | 7189.915587 | 6326.488155 | -9255.95363 | 5850 | 18515.9073 |
| 5 | HomeTeam | 6774.322707 | 5927.101628 | -9048.15719 | 5826 | 18148.3144 |
| 6 | HTAG + logHST | 5991.588813 | 5006.427081 | -8656.79025 | 5849 | 17319.5805 |
| 7 | HTAG + logHC | 7228.580545 | 6382.811467 | -9275.28611 | 5849 | 18556.5722 |
| 8 | HTAG + HR | 7199.906632 | 6346.400575 | -9260.94916 | 5849 | 18527.8983 |
| 9 | HTAG + AR | 7181.604664 | 6321.500495 | -9251.79817 | 5849 | 18509.5963 |
| 10 | HTAG + HomeTeam | 6774.226187 | 5927.150126 | -9048.10893 | 5825 | 18150.2179 |
| 11 | logHST + logHC | 5884.45854 | 4905.019606 | -8603.22511 | 5849 | 17212.4502 |
| 12 | logHST + HR | 5981.051285 | 4999.825831 | -8651.52148 | 5849 | 17309.043 |
| 13 | logHST + AR | 5976.32572 | 4989.730312 | -8649.1587 | 5849 | 17304.3174 |
| 14 | logHST + HomeTeam | 5794.972454 | 4861.503044 | -8558.48207 | 5825 | 17170.9641 |
| 15 | logHC + HR | 7202.165607 | 6345.10234 | -9262.07864 | 5849 | 18530.1573 |
| 16 | logHC + AR | 7184.848102 | 6320.275663 | -9253.41989 | 5849 | 18512.8398 |
| 17 | logHC + HomeTeam | 6772.555481 | 5925.72631 | -9047.27358 | 5825 | 18148.5472 |
| 18 | HR + AR | 7146.205721 | 6274.697767 | -9234.0987 | 5849 | 18474.1974 |
| 19 | HR + HomeTeam | 6744.566014 | 5887.734599 | -9033.27885 | 5825 | 18120.5577 |
| 20 | AR + HomeTeam | 6723.647103 | 5864.579973 | -9022.81939 | 5825 | 18099.6388 |
| 21 | HTAG + logHST + logHC | 5884.276723 | 4905.221604 | -8603.1342 | 5848 | 17214.2684 |
| 22 | HTAG + logHST + HR | 5981.001162 | 4999.831633 | -8651.49642 | 5848 | 17310.9928 |
| 23 | HTAG + logHST + AR | 5976.16023 | 4989.718307 | -8649.07595 | 5848 | 17306.1519 |
| 24 | HTAG + logHST + HomeTeam | 5794.152405 | 4860.793388 | -8558.07204 | 5824 | 17172.1441 |
| 25 | HTAG + logHC + HR | 7195.610658 | 6341.428238 | -9258.80117 | 5848 | 18525.6023 |
| 26 | HTAG + logHC + AR | 7176.774563 | 6315.222943 | -9249.38312 | 5848 | 18506.7662 |
| 27 | HTAG + logHC + HomeTeam | 6772.461085 | 5925.79146 | -9047.22638 | 5824 | 18150.4528 |
| 28 | HTAG + HR + AR | 7140.862976 | 6271.984551 | -9231.42733 | 5848 | 18470.8547 |
| 29 | HTAG + HR + HomeTeam | 6744.534283 | 5887.628172 | -9033.26298 | 5824 | 18122.526 |
| 30 | HTAG + AR + HomeTeam | 6723.629924 | 5864.636893 | -9022.8108 | 5824 | 18101.6216 |
| 31 | logHST + logHC + HR | 5871.014882 | 4896.454709 | -8596.50328 | 5848 | 17201.0066 |
| 32 | logHST + logHC + AR | 5868.2966 | 4889.696454 | -8595.14414 | 5848 | 17198.2883 |
| 33 | logHST + logHC + HomeTeam | 5654.551142 | 4739.117315 | -8488.27141 | 5824 | 17032.5428 |
| 34 | logHST + HR + AR | 5963.481614 | 4981.971641 | -8642.73665 | 5848 | 17293.4733 |
| 35 | logHST + HR + HomeTeam | 5786.463977 | 4854.766729 | -8554.22783 | 5824 | 17164.4557 |
| 36 | logHST + AR + HomeTeam | 5779.462304 | 4843.934353 | -8550.72699 | 5824 | 17157.454 |
| 37 | logHC + HR + AR | 7143.100066 | 6270.844144 | -9232.54587 | 5848 | 18473.0917 |
| 38 | logHC + HR + HomeTeam | 6741.671522 | 5885.079803 | -9031.8316 | 5824 | 18119.6632 |
| 39 | logHC + AR + HomeTeam | 6720.916371 | 5862.947522 | -9021.45402 | 5824 | 18098.908 |
| 40 | HR + AR + HomeTeam | 6688.847894 | 5820.817497 | -9005.41979 | 5824 | 18066.8396 |
| 41 | HTAG + logHST + logHC + HR | 5871.001939 | 4896.516962 | -8596.49681 | 5847 | 17202.9936 |
| 42 | HTAG + logHST + logHC + AR | 5868.190843 | 4889.843636 | -8595.09126 | 5847 | 17200.1825 |
| 43 | HTAG + logHST + logHC + HomeTeam | 5653.260661 | 4737.546129 | -8487.62617 | 5823 | 17033.2523 |
| 44 | HTAG + logHST + HR + AR | 5963.473689 | 4981.971434 | -8642.73268 | 5847 | 17295.4654 |
| 45 | HTAG + logHST + HR + HomeTeam | 5785.105855 | 4853.762116 | -8553.54877 | 5823 | 17165.0975 |
| 46 | HTAG + logHST + AR + HomeTeam | 5778.422587 | 4843.106596 | -8550.20713 | 5823 | 17158.4143 |
| 47 | HTAG + logHC + HR + AR | 7137.855642 | 6268.033498 | -9229.92366 | 5847 | 18469.8473 |
| 48 | HTAG + logHC + HR + HomeTeam | 6741.634064 | 5884.948455 | -9031.81287 | 5823 | 18121.6257 |
| 49 | HTAG + logHC + AR + HomeTeam | 6720.900516 | 5863.011264 | -9021.4461 | 5823 | 18100.8922 |
| 50 | HTAG + HR + AR + HomeTeam | 6688.673371 | 5820.46121 | -9005.33253 | 5823 | 18068.665 |
| 51 | logHST + logHC + HR + AR | 5852.546375 | 4879.897734 | -8587.26903 | 5847 | 17184.5381 |
| 52 | logHST + logHC + HR + HomeTeam | 5643.550689 | 4730.118554 | -8482.77118 | 5823 | 17023.5424 |
| 53 | logHST + logHC + AR + HomeTeam | 5638.062487 | 4723.044811 | -8480.02708 | 5823 | 17018.0542 |
| 54 | logHST + HR + AR + HomeTeam | 5769.202072 | 4835.984595 | -8545.59688 | 5823 | 17149.1938 |
| 55 | logHC + HR + AR + HomeTeam | 6684.566879 | 5817.641422 | -9003.27928 | 5823 | 18064.5586 |
| 56 | HTAG + logHST + logHC + HR + AR | 5852.545665 | 4879.884293 | -8587.26867 | 5846 | 17186.5373 |
| 57 | HTAG + logHST + logHC + HR + HomeTeam | 5641.498093 | 4727.977727 | -8481.74489 | 5822 | 17023.4898 |
| 58 | HTAG + logHST + logHC + AR + HomeTeam | 5636.487104 | 4721.301936 | -8479.23939 | 5822 | 17018.4788 |
| 59 | HTAG + logHST + HR + AR + HomeTeam | 5767.472851 | 4834.830337 | -8544.73226 | 5822 | 17149.4645 |
| 60 | HTAG + logHC + HR + AR + HomeTeam | 6684.374122 | 5817.22362 | -9003.1829 | 5822 | 18066.3658 |
| 61 | logHST + logHC + HR + AR + HomeTeam | 5625.025529 | 4712.701481 | -8473.5086 | 5822 | 17007.0172 |
| 62 | HTAG + logHST + logHC + HR + AR + HomeTeam | 5622.496 | 4710.341455 | -8472.24384 | 5821 | 17006.4877 |

*Bảng 2: Các mô hình được giữ lại sau kiểm định GoodnessOfFit và sắp xếp theo thứ tự AIC tăng dần*

| | model | deviance | pearson_chi2 | llf | df_resid | AIC | p_chisq |
|---|---|---|---|---|---|---|---|
| 0 | HTAG + logHST + logHC + HR + AR + HomeTeam | 5622.496 | 4710.341455 | -8472.24384 | 5821 | 17006.4877 | 0.968193 |
| 1 | logHST + logHC + HR + AR + HomeTeam | 5625.025529 | 4712.701481 | -8473.5086 | 5822 | 17007.0172 | 0.967131 |
| 2 | logHST + logHC + AR + HomeTeam | 5638.062487 | 4723.044811 | -8480.02708 | 5823 | 17018.0542 | 0.957826 |
| 3 | HTAG + logHST + logHC + AR + HomeTeam | 5636.487104 | 4721.301936 | -8479.23939 | 5822 | 17018.4788 | 0.958328 |
| 4 | HTAG + logHST + logHC + HR + HomeTeam | 5641.498093 | 4727.977727 | -8481.74489 | 5822 | 17023.4898 | 0.953925 |
| 5 | logHST + logHC + HR + HomeTeam | 5643.550689 | 4730.118554 | -8482.77118 | 5823 | 17023.5424 | 0.95294 |
| 6 | logHST + logHC + HomeTeam | 5654.551142 | 4739.117315 | -8488.27141 | 5824 | 17032.5428 | 0.942872 |
| 7 | HTAG + logHST + logHC + HomeTeam | 5653.260661 | 4737.546129 | -8487.62617 | 5823 | 17033.2523 | 0.943201 |
| 8 | logHST + HR + AR + HomeTeam | 5769.202072 | 4835.984595 | -8545.59688 | 5823 | 17149.1938 | 0.689303 |
| 9 | HTAG + logHST + HR + AR + HomeTeam | 5767.472851 | 4834.830337 | -8544.73226 | 5822 | 17149.4645 | 0.691715 |
| 10 | logHST + AR + HomeTeam | 5779.462304 | 4843.934353 | -8550.72699 | 5824 | 17157.454 | 0.658196 |
| 11 | HTAG + logHST + AR + HomeTeam | 5778.422587 | 4843.106596 | -8550.20713 | 5823 | 17158.4143 | 0.658345 |
| 12 | logHST + HR + HomeTeam | 5786.463977 | 4854.766729 | -8554.22783 | 5824 | 17164.4557 | 0.633961 |
| 13 | HTAG + logHST + HR + HomeTeam | 5785.105855 | 4853.762116 | -8553.54877 | 5823 | 17165.0975 | 0.635226 |
| 14 | logHST + HomeTeam | 5794.972454 | 4861.503044 | -8558.48207 | 5825 | 17170.9641 | 0.607381 |
| 15 | HTAG + logHST + HomeTeam | 5794.152405 | 4860.793388 | -8558.07204 | 5824 | 17172.1441 | 0.606747 |
| 16 | logHST + logHC + HR + AR | 5852.546375 | 4879.897734 | -8587.26903 | 5847 | 17184.5381 | 0.477098 |
| 17 | HTAG + logHST + logHC + HR + AR | 5852.545665 | 4879.884293 | -8587.26867 | 5846 | 17186.5373 | 0.473419 |
| 18 | logHST + logHC + AR | 5868.2966 | 4889.696454 | -8595.14414 | 5848 | 17198.2883 | 0.423237 |
| 19 | HTAG + logHST + logHC + AR | 5868.190843 | 4889.843636 | -8595.09126 | 5847 | 17200.1825 | 0.420003 |
| 20 | logHST + logHC + HR | 5871.014882 | 4896.454709 | -8596.50328 | 5848 | 17201.0066 | 0.413445 |
| 21 | HTAG + logHST + logHC + HR | 5871.001939 | 4896.516962 | -8596.49681 | 5847 | 17202.9936 | 0.409895 |
| 22 | logHST + logHC | 5884.45854 | 4905.019606 | -8603.22511 | 5849 | 17212.4502 | 0.369438 |
| 23 | HTAG + logHST + logHC | 5884.276723 | 4905.221604 | -8603.1342 | 5848 | 17214.2684 | 0.366587 |
| 24 | logHST + HR + AR | 5963.481614 | 4981.971641 | -8642.73665 | 5848 | 17293.4733 | 0.142982 |
| 25 | HTAG + logHST + HR + AR | 5963.473689 | 4981.971434 | -8642.73268 | 5847 | 17295.4654 | 0.140928 |
| 26 | logHST + AR | 5976.32572 | 4989.730312 | -8649.1587 | 5849 | 17304.3174 | 0.120008 |
| 27 | HTAG + logHST + AR | 5976.16023 | 4989.718307 | -8649.07595 | 5848 | 17306.1519 | 0.118474 |
| 28 | logHST + HR | 5981.051285 | 4999.825831 | -8651.52148 | 5849 | 17309.043 | 0.111612 |
| 29 | HTAG + logHST + HR | 5981.001162 | 4999.831633 | -8651.49642 | 5848 | 17310.9928 | 0.109957 |
| 30 | HTAG + logHST | 5991.588813 | 5006.427081 | -8656.79025 | 5849 | 17319.5805 | 0.094434 |

*Bảng 4: Hệ số của mô hình [1] FTHG ~ HTAG + logHST + logHC + HR + AR + HomeTeam*

| | coef | std err | z | P>|z| | [0.025 | 0.975] |
|---|---|---|---|---|---|---|
| **Intercept** | -0.2387 | 0.062 | -3.881 | 0 | -0.359 | -0.118 |
| **HomeTeam[T.Aston Villa]** | -0.2979 | 0.063 | -4.696 | 0 | -0.422 | -0.174 |
| **HomeTeam[T.Birmingham]** | -0.3592 | 0.102 | -3.535 | 0 | -0.558 | -0.16 |
| **HomeTeam[T.Blackburn]** | -0.3195 | 0.074 | -4.319 | 0 | -0.465 | -0.175 |
| **HomeTeam[T.Bolton]** | -0.4274 | 0.076 | -5.652 | 0 | -0.576 | -0.279 |
| **HomeTeam[T.Bournemouth]** | -0.026 | 0.094 | -0.276 | 0.782 | -0.21 | 0.158 |
| **HomeTeam[T.Chelsea]** | -0.0013 | 0.051 | -0.025 | 0.98 | -0.102 | 0.099 |
| **HomeTeam[T.Crystal Palace]** | -0.277 | 0.087 | -3.181 | 0.001 | -0.448 | -0.106 |
| **HomeTeam[T.Everton]** | -0.1522 | 0.055 | -2.744 | 0.006 | -0.261 | -0.043 |
| **HomeTeam[T.Fulham]** | -0.367 | 0.067 | -5.488 | 0 | -0.498 | -0.236 |
| **HomeTeam[T.Hull]** | -0.3541 | 0.104 | -3.406 | 0.001 | -0.558 | -0.15 |
| **HomeTeam[T.Leeds]** | -0.1812 | 0.098 | -1.84 | 0.066 | -0.374 | 0.012 |
| **HomeTeam[T.Leicester]** | -0.0586 | 0.073 | -0.801 | 0.423 | -0.202 | 0.085 |
| **HomeTeam[T.Liverpool]** | -0.0308 | 0.052 | -0.595 | 0.552 | -0.132 | 0.071 |
| **HomeTeam[T.Man United]** | -0.0041 | 0.051 | -0.079 | 0.937 | -0.104 | 0.096 |
| **HomeTeam[T.Middlesbrough]** | -0.2691 | 0.08 | -3.348 | 0.001 | -0.427 | -0.112 |
| **HomeTeam[T.Newcastle]** | -0.2486 | 0.059 | -4.205 | 0 | -0.365 | -0.133 |
| **HomeTeam[T.Portsmouth]** | -0.5485 | 0.099 | -5.562 | 0 | -0.742 | -0.355 |
| **HomeTeam[T.Southampton]** | -0.169 | 0.067 | -2.518 | 0.012 | -0.301 | -0.037 |
| **HomeTeam[T.Stoke]** | -0.1696 | 0.074 | -2.288 | 0.022 | -0.315 | -0.024 |
| **HomeTeam[T.Sunderland]** | -0.5156 | 0.07 | -7.391 | 0 | -0.652 | -0.379 |
| **HomeTeam[T.Swansea]** | -0.2126 | 0.084 | -2.541 | 0.011 | -0.377 | -0.049 |
| **HomeTeam[T.Tottenham]** | -0.133 | 0.053 | -2.489 | 0.013 | -0.238 | -0.028 |
| **HomeTeam[T.West Brom]** | -0.3478 | 0.071 | -4.912 | 0 | -0.487 | -0.209 |

|   | coef | std err | z | P>|z| | [0.025 | 0.975] |
|---|---|---|---|---|---|---|
| HomeTeam[T.West Ham] | -0.2262 | 0.059 | -3.866 | 0 | -0.341 | -0.112 |
| HomeTeam[T.Wigan] | -0.5976 | 0.085 | -7.011 | 0 | -0.765 | -0.431 |
| HTAG | 0.0246 | 0.015 | 1.597 | 0.11 | -0.006 | 0.055 |
| logHST | 0.6947 | 0.022 | 31.036 | 0 | 0.651 | 0.739 |
| logHC | -0.2298 | 0.019 | -12.265 | 0 | -0.267 | -0.193 |
| HR | -0.1733 | 0.048 | -3.634 | 0 | -0.267 | -0.08 |
| AR | 0.14 | 0.031 | 4.451 | 0 | 0.078 | 0.202 |

*Bảng 6: Hệ số của mô hình [1] sau khi loại bớt quan sát*

|   | coef | std err | z | P>|z| | [0.025 | 0.975] |
|---|---|---|---|---|---|---|
| Intercept | -0.2389 | 0.062 | -3.883 | 0 | -0.359 | -0.118 |
| HomeTeam[T.Aston Villa] | -0.2979 | 0.063 | -4.697 | 0 | -0.422 | -0.174 |
| HomeTeam[T.Birmingham] | -0.3592 | 0.102 | -3.535 | 0 | -0.558 | -0.16 |
| HomeTeam[T.Blackburn] | -0.3193 | 0.074 | -4.316 | 0 | -0.464 | -0.174 |
| HomeTeam[T.Bolton] | -0.4272 | 0.076 | -5.65 | 0 | -0.575 | -0.279 |
| HomeTeam[T.Bournemouth] | -0.0261 | 0.094 | -0.277 | 0.782 | -0.21 | 0.158 |
| HomeTeam[T.Chelsea] | -0.0012 | 0.051 | -0.024 | 0.981 | -0.102 | 0.099 |
| HomeTeam[T.Crystal Palace] | -0.2691 | 0.087 | -3.09 | 0.002 | -0.44 | -0.098 |
| HomeTeam[T.Everton] | -0.1521 | 0.055 | -2.743 | 0.006 | -0.261 | -0.043 |
| HomeTeam[T.Fulham] | -0.3669 | 0.067 | -5.485 | 0 | -0.498 | -0.236 |
| HomeTeam[T.Hull] | -0.3539 | 0.104 | -3.404 | 0.001 | -0.558 | -0.15 |
| HomeTeam[T.Leeds] | -0.1813 | 0.098 | -1.842 | 0.065 | -0.374 | 0.012 |
| HomeTeam[T.Leicester] | -0.0557 | 0.073 | -0.761 | 0.447 | -0.199 | 0.088 |
| HomeTeam[T.Liverpool] | -0.0311 | 0.052 | -0.601 | 0.548 | -0.133 | 0.07 |
| HomeTeam[T.Man United] | -0.0041 | 0.051 | -0.08 | 0.936 | -0.104 | 0.096 |
| HomeTeam[T.Middlesbrough] | -0.2691 | 0.08 | -3.348 | 0.001 | -0.427 | -0.112 |
| HomeTeam[T.Newcastle] | -0.2486 | 0.059 | -4.204 | 0 | -0.364 | -0.133 |
| HomeTeam[T.Portsmouth] | -0.5484 | 0.099 | -5.56 | 0 | -0.742 | -0.355 |
| HomeTeam[T.Southampton] | -0.169 | 0.067 | -2.518 | 0.012 | -0.301 | -0.037 |
| HomeTeam[T.Stoke] | -0.1666 | 0.074 | -2.248 | 0.025 | -0.312 | -0.021 |
| HomeTeam[T.Sunderland] | -0.5155 | 0.07 | -7.389 | 0 | -0.652 | -0.379 |
| HomeTeam[T.Swansea] | -0.2126 | 0.084 | -2.541 | 0.011 | -0.377 | -0.049 |
| HomeTeam[T.Tottenham] | -0.133 | 0.053 | -2.488 | 0.013 | -0.238 | -0.028 |
| HomeTeam[T.West Brom] | -0.3515 | 0.071 | -4.951 | 0 | -0.491 | -0.212 |
| HomeTeam[T.West Ham] | -0.2262 | 0.059 | -3.866 | 0 | -0.341 | -0.112 |
| HomeTeam[T.Wigan] | -0.5912 | 0.085 | -6.936 | 0 | -0.758 | -0.424 |
| HTAG | 0.0246 | 0.015 | 1.602 | 0.109 | -0.006 | 0.055 |
| logHST | 0.6942 | 0.022 | 31.006 | 0 | 0.65 | 0.738 |
| logHC | -0.2291 | 0.019 | -12.221 | 0 | -0.266 | -0.192 |
| HR | -0.1764 | 0.048 | -3.682 | 0 | -0.27 | -0.082 |
| AR | 0.1397 | 0.031 | 4.436 | 0 | 0.078 | 0.201 |

**References**

[1] Alberti, G., Iaia, F.M., Arcelli, E. et al. Goal scoring patterns in major European soccer leagues. Sport Sci Health 9, 151–153 (2013). https://doi.org/10.1007/s11332-013-0154-9.

[2] Lee, Alan. (2012). Modeling Scores in the Premier League: Is Manchester United Really the Best?. CHANCE. 10. 15-19. 10.1080/09332480.1997.10554791.

[3] Tax, Niek & Joustra, Yme. (2015). Predicting The Dutch Football Competition Using Public Data: A Machine Learning Approach. 10.13140/RG.2.1.1383.4729.

[4] https://www.football-data.co.uk/englandm.php

[5] James, G., Witten, D., Hastie, T., & Tibshirani, R. (2021). An Introduction to Statistical Learning: with Applications in R (Springer Texts in Statistics) (2nd ed. 2021 ed.). Springer.

[6] Fox, J., & Weisberg, S. (2018). An R Companion to Applied Regression (3rd ed.). SAGE Publications, Inc.

[7] Fox, J., & Weisberg, S. (2018). An R Companion to Applied Regression (3rd ed.). SAGE Publications, Inc.

[8] Dunn, P. K., & Smyth, G. K. (2018). Generalized Linear Models With Examples in R (Springer Texts in Statistics) (1st ed. 2018 ed.). Springer.

[9] Olsson, U. (2002). Generalized Linear Models: An Applied Approach. Studentlitteratur AB.

[10] C. (2012). Regression Analysis by Example, 5th Edition (5th ed.). Wiley.

[11] Ugarte, M. D., Militino, A. F., & Arnholt, A. T. (2015). Probability and Statistics with R (2nd ed.). Chapman and Hall/CRC.

[12] Nguyễn Văn Tuấn (2006). Ngôn ngữ R trong thống kê suy diễn.

[13] Nguyễn Văn Tuấn (2020). Mô Hình Hồi Quy Và Khám Phá Khoa Học. Nhà Xuất Bản Tổng hợp TP.HCM.